\documentclass[reprint,nofootinbib,amsmath,amssymb,aps,floatfix,superscriptaddress,twocolumn]{revtex4-2}
\usepackage{graphicx}
\usepackage{dcolumn}
\usepackage[colorlinks,linkcolor=magenta,anchorcolor=cyan,citecolor=blue]{hyperref}
\usepackage{bm}
\usepackage[multiple]{footmisc}
\usepackage{lipsum}
\usepackage{soul}

\begin{document}
\title{Thin Accretion Disk Around Rotating Hairy Black Hole: Radiative Property and Optical Appearance}

\author{Zhen Li} 
\email{zhen.li@just.edu.cn}
\affiliation{School of Science, Jiangsu University of Science and Technology, Zhenjiang 212100, China}
\author{Xiao-Kan Guo}
\affiliation{School of Mathematics and Physics, Yancheng Institute of Technology, Jiangsu 224051, China}
\date{\today}

\begin{abstract}
The gravitational decoupling method systematically generates hairy modifications to the solutions in general relativity due to new gravitational sources. In view of the recent advances in astronomical observations, these hairy solutions are expected to be testable in the near-term observations. In this paper, we study the radiative property and optical appearance of the thin accretion disk around the rotating hairy black holes obtained by gravitational decoupling. We numerically compute the radiative flux, temperature, and differential luminosity of the thin accretion disk, and we also show its bolometric image by the ray-tracing method. By comparing with the results for the Kerr metric, we found that the  deviations of the observational properties 
of the thin accretion disk from those of Kerr metric becomes significant in the rapid rotating case, or in the inner region of the disk. These results guide the observational investigations on the rotating hairy black hole.
\end{abstract}
\maketitle

\section{Introduction}
In general relativity (GR), the rotating black holes in the universe are described by the Kerr metric, which is characterized by two parameters: mass and spin. The No Hair Theorem in GR \cite{nh1,nh2,nh3,nh4,nh5} asserts that, apart from mass and spin, no additional distinguishing parameters or "hairs" exist for rotating black holes. However, the discovery of dark matter and dark energy, which might arise from novel fundamental fields, suggests the possibility that black holes could possess extra hair beyond the mass and spin. It has been shown that in the case of a scalar field minimally coupled with GR, certain scalar field potentials can violate the weak energy condition, thereby violating the no-hair theorem and giving rise to hairy black holes \cite{ch1,ch2,ch3,ch4}. These additional hairs could lead to deviations from the GR solutions, like the Kerr metric \cite{dk}. Building on this idea, a new class of hairy black hole, as well as its rotating counterpart, have been proposed using the gravitational decoupling approach \cite{gd0,gd}, a method designated to account for modifications to GR solutions due to extra unknown fields or matter sources \cite{gd1,gd2}. This class of hairy black holes has prompted a wealth of both theoretical and observational investigations \cite{inv1,inv2,inv20,inv21,inv3,inv30,inv31,inv4,inv40, inv41, inv5,inv6,LY,inv7}.

Recent advances in astronomical techniques have greatly improved the accuracy of accretion disk observations \cite{ao1, ao2, ao3, ao4}, enabling more precise measurements of key observables such as temperature and luminosity. Furthermore, the Event Horizon Telescope has achieved the remarkable milestone of capturing black hole images \cite{bh1, bh2}, which inherently reveals the image of the accretion disks \cite{lk1, lk2}. These breakthroughs provide invaluable empirical data for testing black hole models beyond GR. As such, investigating the effects of modified black hole models on the accretion disk in comparison with observational data becomes
 feasible. Such studies could shed light on the fundamental nature of black holes and their behaviors.

The thin accretion disk model, initially proposed by Novikov and Thorne \cite{t1}, Page and Thorne \cite{t2}, and Shakura and Sunyaev \cite{t3}, assumes that the disk is both geometrically thin and optically thick. Under these conditions, the disk emits radiation that closely approximates blackbody emission at each radius. This approximation significantly simplifies the mathematical treatment of the disk's behavior, making it a widely used framework for studying accretion processes around black holes. Despite being formulated decades ago, this model continues to be influential, with numerous studies examining the observational features of the thin accretion disks in various black hole spacetimes \cite{ts0, ts1,ts2,ts3,ts4,ts5,ts6,ts7,ts8,ts9,ts10,ts11,ts12,ts13, ts14,ts15,ts16,ts17}. 

Several studies have explored the imaging of hairy black holes based on empirical intensity profiles, though these may not align with conventional accretion disk models. Most of these works focus on either non-rotating hairy black holes \cite{inv30,inv31,inv4,inv40,inv41} or only the shadow of rotating hairy black holes \cite{inv1,inv2,inv20,inv21}. Given that real astronomical bodies possess angular momentum, the rotating black hole model is more physically realistic than the spherical one. 
But the studies on the accretion disk of rotating hairy black hole remains insufficient. This study aims to fill this gap by investigating the observational characteristics of a thin accretion disk around a rotating hairy black hole. We will examine how variations in the hairy parameters influence the accretion disk's 
radiation properties and its optical appearances, including the bolometric image, thus laying the groundwork for future studies that aim to test alternative hairy black hole models through observations.

The structure of this paper is as follows: In Sec.\ref{sec2}, we introduce the rotating hairy black hole and explore key quantities of time-like and null-like geodesics relevant for analyzing the observational features. In Sec.\ref{sec3}, we investigate the emission properties of thin accretion disk in this spacetime, focusing on how the hairy parameters influence the disk's observables. Sec.\ref{sec4} presents the realistic bolometric images of thin accretion disk using the ray-tracing method, and compares these results with those of the Kerr black hole, highlighting the impact of the rotating hairy black hole on lensing features, redshift distributions and the images. Finally, Sec.\ref{sec5} concludes. Geometrized units with $G = c = 1$ are used throughout this work.

\section{Geodesics of rotating hairy black hole}\label{sec2}

In the Boyer-Lindquist coordinates $(t, r, \theta, \phi)$, the metric of a rotating hairy black hole can be expressed in a specific form as follows \cite{gd},
\begin{equation}\label{metric}
d s^{2}=g_{t t} d t^{2}+g_{r r} d r^{2}+g_{\theta \theta} d \theta^{2}+g_{\phi \phi} d \phi^{2}+2 g_{t \phi} d t d \phi
\end{equation}
where the metric components are given by
\begin{align} \label{mc}
g_{t t}&= -\left[\frac{\Delta-a^{2} \sin ^{2} \theta}{\Sigma}\right], \quad \quad\quad\quad\quad \quad\,\,\,
g_{r r} =\frac{\Sigma}{\Delta}, \nonumber\\
g_{t\phi}&=- a \sin ^{2} \theta\left[1-\frac{\Delta-a^{2} \sin ^{2} \theta}{\Sigma}\right],\quad\quad g_{\theta \theta} =\Sigma, \nonumber\\
g_{\phi\phi}&=\sin ^{2} \theta\left[\Sigma+a^{2} \sin ^{2} \theta\left(2-\frac{\Delta-a^{2} \sin ^{2} \theta}{\Sigma}\right)\right],
\end{align}
with $\Delta=r^{2}+a^{2}-2 M r+\delta r^{2} e^{-r /\left(M-\frac{h_{0}}{2}\right)}$, and $\Sigma=r^{2}+a^{2} \cos ^{2} \theta$. The black hole mass and spin are represented by $M$ and $a$, respectively. The primary hair, denoted as $h_0$, quantifies the entropy produced by the black hole's hair and must satisfy the condition $h_0 \le 2M \equiv h_K$ in order to achieve asymptotic flatness. The parameter $\delta$ describes the deviation from the standard Kerr black hole, and it is related to $h_0$ through the relation $h_0 = \lambda h$. When $\delta = 0$, the spacetime reduces to the Kerr metric. By numerically solving $\Delta =0$, we could obtain the event horizon $r_+$. We plot the $r_+$ as a function of spin, with different hairy parameters $\delta$ and $h_0$, as shown in Fig.~\ref{rms}. All $r_+$ exhibit a decrease as the black hole spin increases. The larger the value of $\delta$ and the smaller the value of 
$h_0$, the faster the decrease. 

The motion of test particles or photons in this hairy black hole spacetime, similar to the Kerr black hole, is governed by three constants of motion: energy $E$, angular momentum along the spin axis $L$, and the Carter constant $Q$ \cite{geo}. The latter two quantities can be alternatively expressed as energy-scaled quantities, namely $\lambda \equiv L/E$ and $\eta \equiv Q/E$. The geodesic equations for a test particle or photon in the spacetime described by metric (\ref{metric}) are then given by the corresponding forms,
\begin{align}
\frac{\Sigma}{E} p^{r}&= \pm_{r} \sqrt{\mathcal{R}(r)}, \label{pr}\\
\frac{\Sigma}{E} p^{\theta} &= \pm_{\theta} \sqrt{\Theta(\theta)},  \label{p8}\\
\frac{\Sigma}{E} p^{\phi}&=\frac{a}{\Delta}\left(r^{2}+a^{2}-a \lambda\right)+\frac{\lambda}{\sin ^{2} \theta}-a , \label{pp}\\
\frac{\Sigma}{E} p^{t}&=\frac{r^{2}+a^{2}}{\Delta}\left(r^{2}+a^{2}-a \lambda\right)+a\left(\lambda-a \sin ^{2} \theta\right) ,\label{pt}
\end{align}
where the symbols $\pm_r$ and $\pm_\theta$ indicate the signs of the radial and angular components of the momentum, $p^r$ and $p^\theta$, respectively. The functions $\mathcal{R}(r)$ and $\Theta(\theta)$ represent the radial and angular potentials and are defined as,
\begin{align}
\mathcal{R}(r)&=\left(r^{2}+a^{2}-a \lambda\right)^{2}-\Delta\left[\mu^2 r^2+\eta+(\lambda-a)^{2}\right], \\
\Theta(\theta)&=\eta+a^{2} \cos ^{2} \theta-\lambda^{2} \cot ^{2} \theta .
\end{align}
For photons, the value of $\mu$ is set to 0. In contrast, for massive test particles, we can set $\mu = 1$, which implies that the conserved quantities $E$, $\lambda$, and $\eta$ are normalized with respect to the particle's mass $\mu$. This means that we are examining the conserved quantities of timelike geodesics in the per unit mass basis.

\subsection{Key quantities of time-like geodesics}

To study the observational properties of thin accretion disks, we begin by considering the particles that make up the disk. These particles are assumed to be massive and follow circular orbits in the equatorial plane, moving along time-like geodesics. So our primary interest is in the particles moving in circular orbits at the equatorial plane, where $Q=0$ and $\theta = \pi/2$. However, we still present a more general form for the key quantities. This general expression  offers a broader applicability beyond the specific case of the equatorial plane.

The angular velocity of a test particle can be derived from the geodesic equations (\ref{pr}-\ref{pt}) and is expressed in terms of the metric components (\ref{mc}) as
\begin{equation}\label{omega}
\Omega=\frac{-\partial_{r} g_{t \phi} \pm \sqrt{\left(\partial_{r} g_{t \phi}\right)^{2}-\left(\partial_{r} g_{t t}\right)\left(\partial_{r} g_{\phi \phi}\right)}}{\partial_{r} g_{\phi \phi}}.
\end{equation}
By utilizing the expression for $\Omega$, we can derive the energy and angular momentum per unit mass of a test particle as follows,
\begin{equation}\label{en}
E = -\frac{g_{t t} + g_{t \phi} \Omega}{\sqrt{-g_{t t} - 2 g_{t \phi} \Omega - g_{\phi \phi} \Omega^{2}}},
\end{equation}
\begin{equation}\label{an}
L = \frac{g_{t \phi} + g_{\phi \phi} \Omega}{\sqrt{-g_{t t} - 2 g_{t \phi} \Omega - g_{\phi \phi} \Omega^{2}}} ,
\end{equation}
where we set $\mu =1$ for simplicity. Another important quantity to consider is the time component of the four-velocity of the test particle, which is expressed as 
\begin{equation}\label{ut}
u^{t} = \frac{1}{\sqrt{-g_{t t}-2 \Omega g_{t \varphi}-\Omega^{2} g_{\varphi \varphi}}}.
\end{equation}
The final quantity to consider is the radius of the innermost stable circular orbit (ISCO), denoted as $r_{\text{ISCO}}$, for the test particle. It can be determined by the vanishing of the second radial derivative of the radial potential,
\begin{equation}
\frac{d^2\mathcal{R}(r)}{dr^2}\bigg| _{r_{\text{ISCO}}} \,= 0 .
\end{equation}
Unlike the Kerr metric, there is no analytical expression for ISCO radius $r_{\text{ISCO}}$ in rotating hairy black hole. Instead, we compute this quantity numerically. The numerical method we employ follows the approach outlined in \cite{che}.

\begin{figure}[t]
  \centering
    \includegraphics[scale = 0.52]{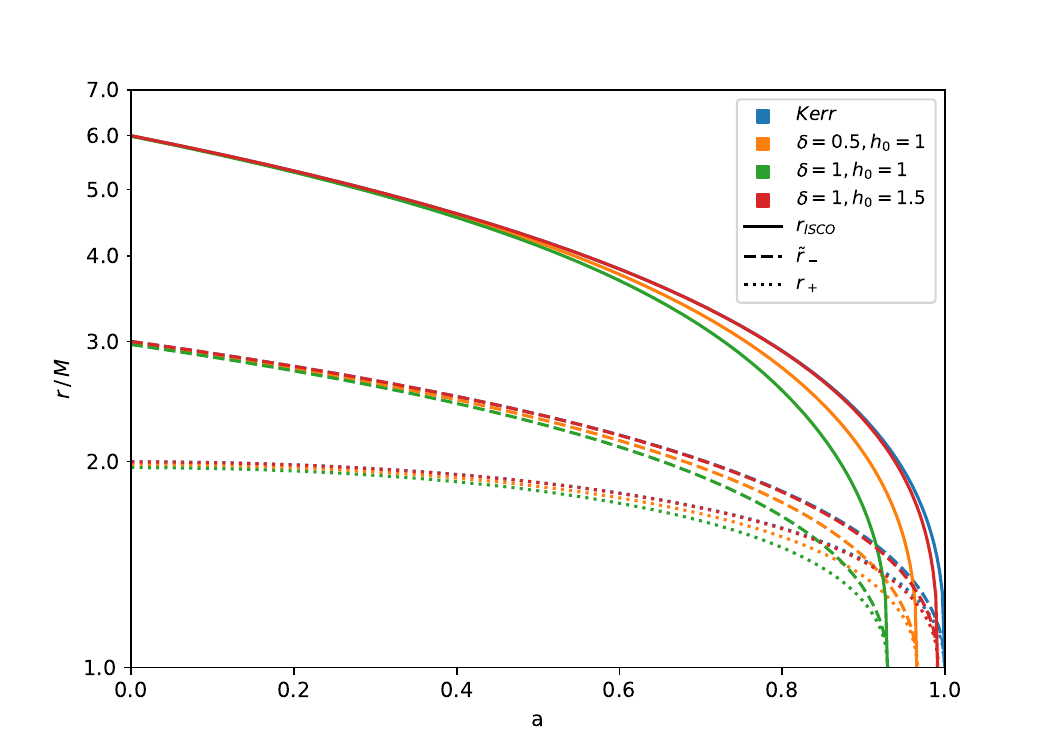}
\caption{The dependence of various radii on the spin parameter $a$ and the hairy parameters $(\delta, h_0)$. The plot shows the behavior of the innermost stable circular orbit radius $r_{\text{ISCO}}$, the inner photon shell radius $\tilde{r}_- $, and the event horizon radius $r_+$. }
\label{rms}
\end{figure}

The values of $r_{\text{ISCO}}$ with different combinations of the hairy parameters ($\delta, h_0$) as a function of black hole spin $a$, are presented in Fig.~\ref{rms}. We observe that as the spin of the black hole increases, the radius $r_{\text{ISCO}}$ decreases. This continues until the spin reaches a critical maximum value, at which $r_{\text{ISCO}}$ coincides with the event horizon $r_+$. Beyond this critical spin value, no horizon exists, and the black hole becomes a naked singularity. Notably, the larger the value of $\delta$ and the smaller the value of $h_0$, the faster the decrease in $r_{\text{ISCO}}$, and the smaller the critical spin. These two parameters, $\delta$ and $h_0$, appear to play opposite roles in the geometry of the rotating hairy black hole. 
\subsection{Key radii of null geodesics}
For photons, with $\mu = 0$, they follow null geodesics. While photons are not directly related to the composition of a thin accretion disk, they play a critical role in the ray-tracing method, which is essential for generating black hole images. In this context, we will focus on several key radii that are frequently used in the ray-tracing calculations.

The roots of the radial potential for a rotating hairy black hole categorize photon trajectories into distinct types, such as those that fall into the black hole or those that scatter back to infinity. These roots govern the behavior of photons and, consequently, have a direct impact on the imaging results of the rotating hairy black hole. To determine these roots, one must solve the equation $\mathcal{R}(r) = 0$, or equivalently,
\begin{equation}\label{14}
r^{4}+\mathcal{A} r^{2}+\mathcal{B} r+\mathcal{C}=0,
\end{equation}
where
\begin{equation}
\begin{aligned}
\mathcal{A} & =a^{2}-\eta-\lambda^{2} ,\\
\mathcal{B} & =2(M-\delta\frac{r}{2} e^{-r/(M-\frac{h_0}{2})})
\left[\eta+(\lambda-a)^{2}\right],\\
\mathcal{C} & =-a^{2} \eta .
\end{aligned}
\end{equation}
We have expressed the equation \eqref{14} in a quadratic form for comparison with the results in the Kerr metric. However, due to the presence of hairy terms, the equation is no longer quadratic, as  $\mathcal{B}$ is dependent on $r$. In fact, the equation \eqref{14} has more than four roots, and solving it analytically is not feasible. Fortunately, most of the additional roots lie deep inside the horizon and appear as complex conjugates. This allows us to focus on the largest four roots, denoted as $r_1, r_2, r_3, r_4$, where the real parts of the roots satisfy $\Re (r_1) < \Re (r_2) < \Re (r_3) < \Re (r_4) $. Consequently, the classification of photon trajectories remains similar to that in the Kerr black hole \cite{lk1, lk2}.

We aim to determine the critical radius $\tilde{r}_{\pm}$, which defines the boundary of photon shells or bounded photon orbits in the spacetime of a rotating hairy black hole. The radii of these bounded photon orbits, denoted as $\tilde{r}$, can be derived by solving the following equation,
\begin{equation}
\mathcal{R}(r)\bigg| _{\tilde{r}} =\frac{d\mathcal{R}(r)}{dr}\bigg| _{\tilde{r}}=0.
\end{equation}
The conserved quantities $\lambda$ and $\eta$ can only assume the following distinct values for $\tilde{r}>r_+$ in order to satisfy the equation above,
\begin{align}
\tilde{\lambda}&=a+\frac{\tilde{r}}{a}\left[\tilde{r}-\frac{2 \tilde{\Delta}}{\tilde{r}-M-N\tilde{r}}\right], \label{cr1}\\
\tilde{\eta}&=\frac{\tilde{r}^{3}}{a^{2}}\left[\frac{4 \tilde{\Delta}(M -N\tilde{r})}{(\tilde{r}-M - N\tilde{r})^{2}}-\tilde{r}\right],\label{cr2}\\
N &= -\frac{\delta e^{2 \tilde{r}/(-2+h_0)}(-2+h_0+2 \tilde{r})}{2(-2+h_0)} ,\label{cr3}
\end{align}
where $\tilde{\Delta} = \tilde{r}^{2}+a^{2}-2 M \tilde{r}+\delta \tilde{r}^{2} e^{-\tilde{r} /\left(M-\frac{h_{0}}{2}\right)}$. In the equatorial plane, with the condition $\tilde{\eta} = 0$, two radii emerge that are greater than the horizon: $\tilde{r}_\pm > r_+$, where $\tilde{r}_-$ and $\tilde{r}_+$ represent the inner and outer boundaries of the photon shell, respectively. The radii of bounded orbits, denoted as $\tilde{r}$, are confined within the range $\tilde{r} \in [\tilde{r}_-, \tilde{r}_+]$. These bounded orbits are unstable and define a critical curve in the image plane of the observer, as discussed in Sec.~\ref{sec4}.

In Fig.~{\ref{rms}}, we present the inner edge of the photon shell, $\tilde{r}_-$, for various values of the black hole spin $a$ and the hairy parameters $(\delta, h_0)$. The behavior of $\tilde{r}_-$ is similar to that of $r_{\text{ISCO}}$. As the spin of the black hole increases, the radius $\tilde{r}_-$ decreases. This reduction continues until the spin reaches a critical value, at which $\tilde{r}_-$ coincides with the event horizon $r_+$ and $r_{\text{ISCO}}$. The larger the value of $\delta$ and the smaller the value of $h_0$, the more rapidly $\tilde{r}_-$ decreases, and the smaller the critical spin.

\section{Radiative properties of thin accretion disks}\label{sec3}

\begin{figure}[t]
  \centering
    \includegraphics[scale = 0.52]{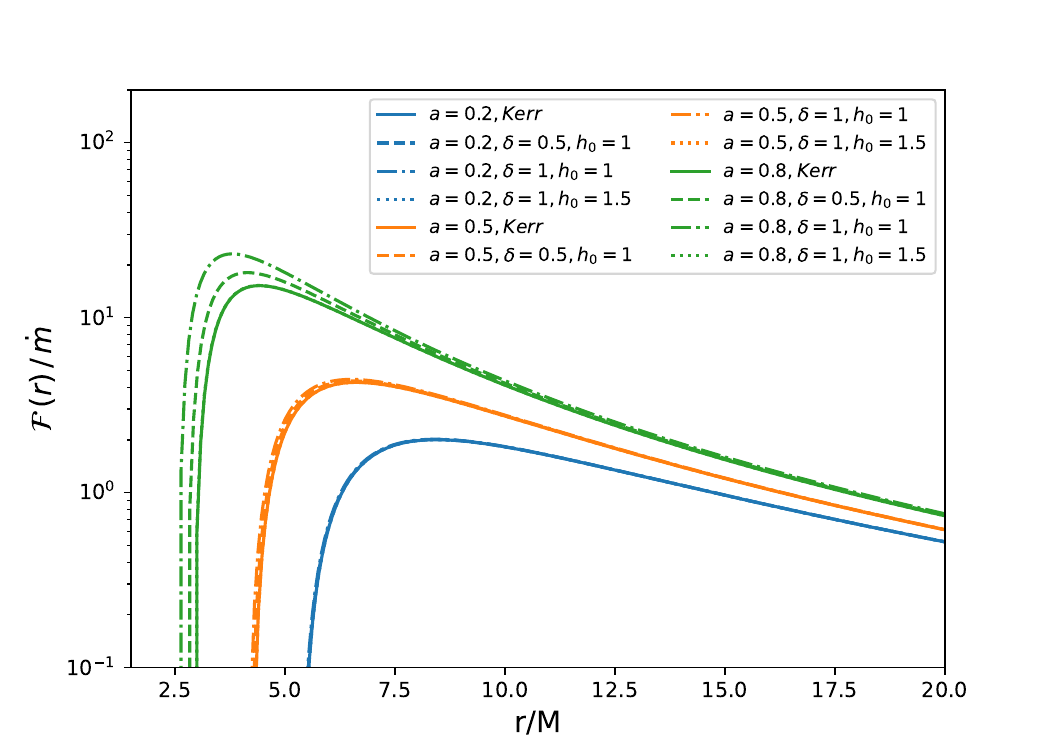}
\caption{The radiative flux per accretion rate of thin accretion disk $\mathcal{F}(r)/\dot m$ as a function of $r/M$ for different black hole spin $a$ and hairy parameters $(\delta, h_0)$. The results of Kerr black hole are also provided as references. All the values are multiplied by $10^5$ for convenience.}
\label{f}
\end{figure}

In the context of the thin accretion disk model \cite{t1,t2,t3}, the radiative flux $\mathcal{F}(r)$ characterizes the energy emitted per unit area and per unit time at radius $r$ from the disk. This flux can be expressed as
\begin{equation}\label{flux}
\mathcal{F}(r)=-\frac{\dot m }{4 \pi \sqrt{-g}} \frac{\Omega_{, r}}{(E-\Omega L)^{2}} \int_{r_{I S C O}}^{r}(E-\Omega L) L_{, \tilde{r}} d \tilde{r},
\end{equation}
where $\dot m $ is the mass accretion rate. We consider that the mass accretion rate is much smaller than the Eddington accretion rate, such that the thin accretion disk model holds. Additionally, $g$ denotes the determinant of the three-dimensional subspace spanned by the coordinates $(t, r, \phi)$.
\begin{equation}
g = g_{rr}(g_{tt}g_{\phi\phi}-g_{t\phi}^2).
\end{equation}
$\Omega$, $E$, and $L$ are given by Eq.~(\ref{omega}), (\ref{en}) and (\ref{an}), respectively. The radial derivatives of angular velocity and angular momentum for a test particle are denoted as $\Omega_{,r}$ and $L_{,r}$, respectively. In this analysis, we focus on a thin accretion disk confined to the equatorial plane, where $\theta = \pi/2$. This assumption significantly simplifies both the metric components and thus the expressions for the angular velocity $\Omega$, energy $E$, and angular momentum $L$. In Fig.~\ref{f}, we present the dependence of the radiative flux on the radius for various values of the black hole spin $a$ and hairy parameters $(\delta = 0.5, h_0=1)$, $(\delta = 1, h_0=1)$, and $(\delta = 1, h_0=1.5)$. For convenience, the values shown in the figure are scaled by a factor of $10^5$.

We observe that the radiative flux initially increases and then decreases with radius. In the far regions, the differences between the curves are small. However, significant differences are observed in the inner region of the thin accretion disk. For slowly rotating hairy black holes, the radiative flux is almost the same as that of the Kerr black hole. For rapidly rotating black holes, the differences between the Kerr black hole and the rotating hairy black hole become more pronounced. The larger the spin, the greater the deviations, especially in the inner region of the thin accretion disk. In addition, the radius where the radiative flux drops to zero, or the inner edge of the disk, is $r_{\text{ISCO}}$, and they also decrease with spin as we discussed in the above section. A higher spin could eliminate the degeneracy between the rotating hairy black hole and the Kerr black hole, suggesting that observations should focus more on rapidly rotating black holes, as they could help us distinguish between different black hole spacetimes. Additionally, the larger the value of $\delta$ and the smaller the value of $h_0$, the greater the radiative flux values, indicating a higher radiative efficiency.

\begin{figure}[t]
  \centering
    \includegraphics[scale = 0.52]{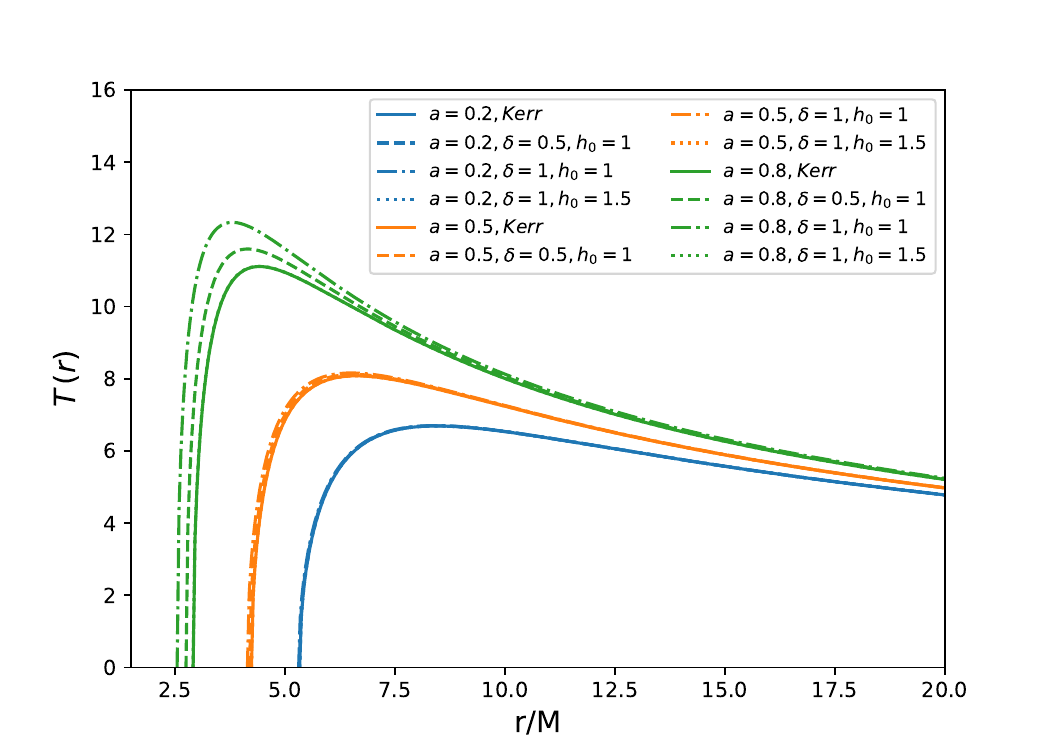}
\caption{The temperature of thin accretion disk $T(r)$ as a function of $r/M$ for different black hole spin $a$ and hairy parameters $(\delta, h_0)$. We set the Stefan–Boltzmann constant $\sigma =1$, and the values in this figure are multiplied by $10^2$ for convenience}
\label{t}
\end{figure}

Assuming that the thin accretion disk is in thermal equilibrium at each radius $r$, the temperature can be derived using the Stefan-Boltzmann law.
\begin{equation}
T = \sqrt[4]{\mathcal{F}(r)/\sigma},
\end{equation}
where $\sigma$ is the Stefan-Boltzmann constant. Fig.~\ref{t} presents the temperature distribution as a function of radial distance for various black hole spins $a$ and hairy parameters $(\delta, h_0)$. For comparison, the results for the Kerr black hole are also included. In this figure, the Stefan-Boltzmann constant $\sigma$ is set to 1, and all values are scaled by a factor of $10^2$ for ease of presentation.

The overall behavior of the temperature follows that of the radiative flux, as the temperature is simply the quartic root of the latter. From Fig.~\ref{t}, it is evident that the temperature initially increases and then decreases with radius. The most significant differences are observed in the inner regions of the accretion disk. The rotating hairy black holes are hotter than the Kerr black hole, especially in the case with hairy parameters $(\delta = 1, h_0 = 1)$. Additionally, the temperature of the thin accretion disk increases as the spin increases. Larger values of $\delta$ and smaller values of $h_0$ further enhancing the disk temperature. This means that the thin accretion disk is more easily detectable in observationas, as the increased temperature would lead to stronger emission and higher luminosity.

\begin{figure}[t]
  \centering
    \includegraphics[scale = 0.52]{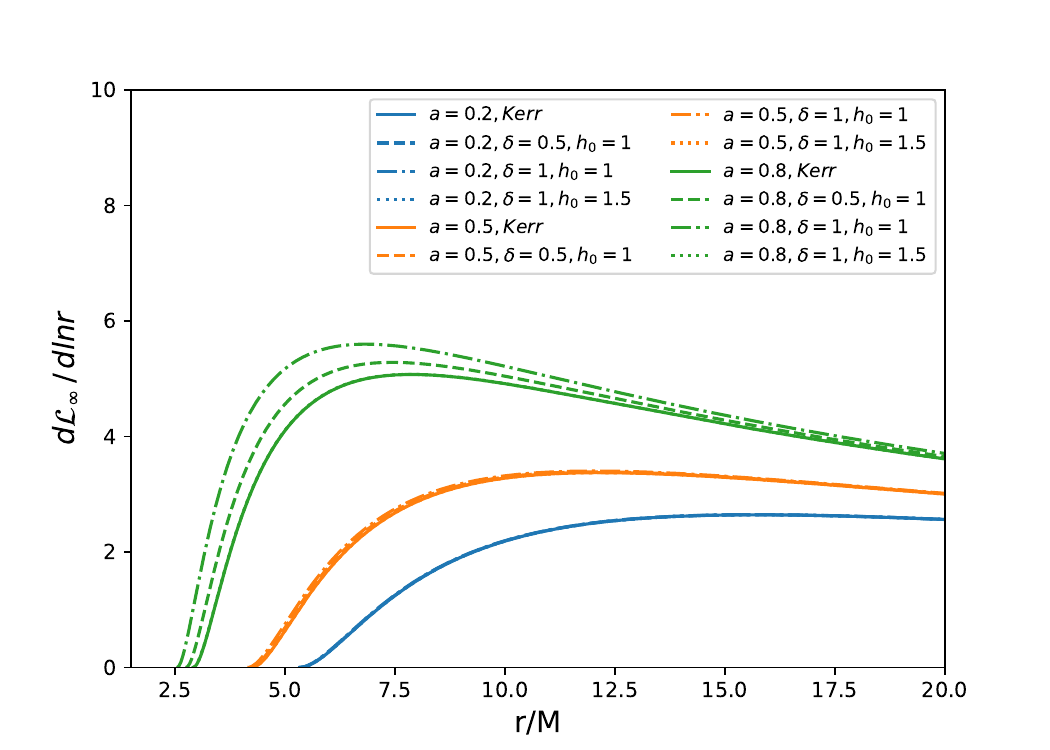}
\caption{Differential luminosity of thin accretion disk $d\mathcal{L}_{\infty}/d\ln r$ as a function of $r/M$ for different black hole spin $a$ and hairy parameters $(\delta, h_0)$. The results of Kerr black hole are also provided as references. The values in this figure are multiplied by $10^2$.}
\label{d}
\end{figure}

So, we shall explore the differential luminosity of a thin accretion disk, a parameter that can be directly measured through observations. This quantity is expressed as,
\begin{equation}
\frac{d \mathcal{L}_{\infty}}{d \ln r}=4 \pi r \sqrt{-g} E \mathcal{F}(r),
\end{equation}
where $\mathcal{L}_{\infty}$ represents the energy per unit of time reaching an observer at infinity. In Fig.~\ref{d}, we present the differential luminosity as a function of radial distance for various values of black hole spin parameter $a$ and hairy parameters $(\delta, h_0)$. For clarity, the values displayed in this figure have been scaled by a factor of $10^2$.

The overall behavior of the differential luminosity as a function of radius closely resembles that of the radiative flux and temperature. However, there is a notable difference: for both slowly rotating hairy black holes and Kerr black holes, the differential luminosity monotonically increases with the radius. As the spin increases, this monotonic increase disappears, and instead, the luminosity first increases and then decreases with radius. The rotating hairy black holes tend to enhance the luminosity, especially for those with rapid rotation. Additionally, the differential luminosity is found to be larger for rotating hairy black holes with a higher value of $\delta$ and a smaller $h_0$.

\section{The bolometric image of thin accretion disk}\label{sec4}

\subsection{Ray-tracing method}
 
To obtain the black hole image, we employ the ray-tracing technique. This method involves tracing light rays from the observer's image plane back toward the vicinity of the black hole. By determining how the light rays interact with the emission source or the thin accretion disk, we can calculate the corresponding observed flux in the image plane.

The position of a photon on the image plane, denoted by $(\alpha, \beta)$, is determined by its conserved quantities $(\lambda, \eta)$ \cite{ab}. If an observer is located at a radial distance $r_o$ from the center of the black hole, with an inclination angle $\theta_o$, the following relation can be established \cite{ab}, 
\begin{equation}
\alpha=-\frac{\lambda}{\sin \theta_{o}}, \quad \beta = \pm_{o} \sqrt{\eta+a^{2} \cos ^{2} \theta_{o}-\lambda^{2} \cot ^{2} \theta_{o}},
\end{equation}
where $\pm_{o}$ is the sign of $\cos(\theta_{o})$, reversely
\begin{equation}\label{le}
\lambda=-\alpha \sin \theta_{o},\quad \eta=\left(\alpha^{2}-a^{2}\right) \cos ^{2} \theta_{0}+\beta^{2}.
\end{equation}
The conserved quantities associated with photon shells or bounded orbits, denoted by $(\tilde{\lambda}, \tilde{\eta})$ (as given in Eq.~\ref{cr1}-\ref{cr3}), determine the critical curve $\mathcal{C} = (\tilde{\alpha}, \tilde{\beta} )$ in the image plane. This curve is significant as it marks the boundary of the black hole shadow, which corresponds to the region enclosed by the critical curve. The critical curve divides the behavior of light rays when traced backward from the image plane to the rotating hairy black hole. Light rays inside this critical curve will eventually fall into the black hole, while rays outside the curve will be scattered back toward infinity. This indicates that the light rays outside the critical curve have a turning point in their trajectory. The radius of this turning point corresponds to the largest root of the radial potential, $r_4$, which lies beyond the event horizon.

The relation given in Eq.~(\ref{le}) alone is insufficient to uniquely determine the source coordinates of a photon. Specifically, for a black hole with spin $a$, an observer located at a radial coordinate $r_o$ and inclination angle $\theta_o$, the objective is to establish the connection between the coordinates of the image plane $(\alpha, \beta)$ and the source coordinates $(r_s, \theta_s, \phi_s, t_s)$, from which the photon originates. To achieve this, we must parameterize the photon’s trajectory by solving the geodesic equations (\ref{pr}-\ref{pt}). As discussed in \cite{lk1,lk2}, the differential form of the geodesic equation in the spacetime of a rotating hairy black hole can be reformulated into an integral equation,
\begin{align}
I_{r} & =G_{\theta}, \label{ig} \\
\Delta \phi \equiv \phi_{o}-\phi_{s} & =I_{\phi}+\lambda G_{\phi}, \label{dp}\\
\Delta t \equiv t_{o}-t_{s} & =I_{t}+a^{2} G_{t},\label{dt}
\end{align}
where the radial integrals
\begin{align}
I_{r} & =\int_{r_{s}}^{r_{o}}\hspace{-1.85em}- ~\frac{\mathrm{~d} r}{ \pm_{r} \sqrt{\mathcal{R}(r)}},\label{ir}\\
I_{\phi}  &=\int_{r_{s}}^{r_{o}}\hspace{-1.85em}-~~ \frac{a(2 Mr-\delta r^2 e^{-r/(M-\frac{h_0}{2})}-a \lambda)}{ \pm_{r} \Delta(r) \sqrt{\mathcal{R}(r)}} \mathrm{d} r, \label{ip}\\
I_{t} & =\int_{r_{s}}^{r_{o}}\hspace{-1.85em}-~~ \frac{r^{2} \Delta(r)+2 Mr \left(r^{2}+a^{2}-a \lambda\right)}{ \pm_{r} \Delta(r) \sqrt{\mathcal{R}(r)}} \mathrm{d}r \nonumber\\
&+ \int_{r_{s}}^{r_{o}}\hspace{-1.85em}-~~ \frac{(-\delta r^2 e^{-r/(M-\frac{h_0}{2})}) \left(r^{2}+a^{2}-a \lambda\right)}{ \pm_{r} \Delta(r) \sqrt{\mathcal{R}(r)}} \mathrm{d}r
, \label{it}  
\end{align}
and the angular integrals
\begin{align}
G_{\theta} & =\int_{\theta_{s}}^{\theta_{o}}\hspace{-1.85em}- ~~ \frac{\mathrm{~d} \theta}{ \pm_{\theta} \sqrt{\Theta(\theta)}},\label{g8}\\
G_{\phi} &=\int_{\theta_{s}}^{\theta_{o}}\hspace{-1.85em}- ~~ \frac{\csc ^{2} \theta}{ \pm_{\theta} \sqrt{\Theta(\theta)}} \mathrm{d} \theta, \label{gp}\\
G_{t} & =\int_{\theta_{s}}^{\theta_{o}}\hspace{-1.85em}- ~~ \frac{\cos ^{2} \theta}{ \pm_{\theta} \sqrt{\Theta(\theta)}} \mathrm{d} \theta,\label{gt}
\end{align}
Here, the notation $\int\hspace{-1em}-$ represents the (path) integrals along the photon trajectory, which increase monotonically as the photon moves along its path. It is evident that the angular integrals for a rotating hairy black hole are identical to those in the Kerr black hole \cite{lk1,lk2}.

For the Kerr black hole, the radial integrals can be computed analytically and are invertible. In contrast, for a rotating hairy black hole, the radial integrals are not expressible in closed form and must be computed numerically. Nevertheless, by inverting these integrals, we can determine the source coordinates corresponding to a photon with known conservative quantities $(\lambda, \eta)$.

Once the source coordinates are determined, the corresponding flux at this spacetime point can be computed based on the flux profile. Photons are then added to the image plane at the position $(\alpha, \beta)$, ultimately producing the black hole's bolometric image. In this study, we focus on a stationary axisymmetric thin accretion disk, where the emission profile depends only on the radius $r_s$, see Eq. (\ref{flux}). A source ring at radius $r_s$ in the equatorial plane (the thin accretion disk) may produce multiple images in the observer's plane, as photons can be lensed multiple times by the rotating hairy black hole. Photons that cross the equatorial plane $n$ times along their trajectory between the source and the observer contribute to the $n$-th lensing band in the image plane. For instance, the $n=0$ band corresponds to direct images without any equatorial crossings. Photons that undergo one equatorial crossing belong to the $n=1$ lensing band. Higher values of $n$ correspond to narrower and fainter lensing bands, located closer to the critical curve \cite{lk1,lk2}.

To obtain the composite image of the black hole, we need to sum the contributions from all lensing bands. However, for simplicity, we limit our consideration up to the first two lensing bands, as they nearly coincide with the critical curve and the higher bands contribute much less to the observed image. By utilizing the radiative flux profile at the source coordinates, we can calculate the observed flux in the image plane using the following relation,
\begin{equation}\label{oflux}
\mathcal{F}_{0}(\alpha, \beta)=\sum_{n=0}^{2}  \chi ^{4}\left(r_{\mathrm{s}}^{(n)}, \alpha, \beta\right) \mathcal{F}_{\mathrm{s}}\left(r_{\mathrm{s}}^{(n)}\right),
\end{equation}
here $\chi $ is the redshift factor and takes the form as
\begin{equation}\label{red}
\chi \equiv \frac{1}{1+z} = \frac{1}{u^t(1-\lambda \Omega)}.
\end{equation}
where $z$ is the redshift parameter of a photon.

\subsection{Images of rotating hairy black hole}

Using the ray-tracing method described above, we modified the well-known \textbf{aart}\footnote{\url{https://github.com/iAART/aart}} code \cite{aart} to enable the numerical computation of observables, thus extending its applicability to non-Kerr black hole models. We examine the lensing features, redshift distributions, and bolometric images of rotating hairy black holes as a function of black hole spin, the hairy parameters, and the inclination angle. Additionally, we conduct a comparative analysis between rotating hairy black hole and Kerr black hole, emphasizing the distinctions introduced by the hairy parameters.

As discussed in Sec.~\ref{sec2} and Sec.~\ref{sec3}, the rotating hairy black hole with hairy parameters $(\delta = 1, h_0 = 1)$ exhibits more pronounced and significant differences when compared to the Kerr black hole. It also captures the general trends observed in other hairy parameter sets $(\delta = 0.5, h_0 = 1)$ and $(\delta = 1, h_0 = 1.5)$. Therefore, in this section, we will focus on comparing the rotating hairy black hole with parameters $(\delta = 1, h_0 = 1)$ to the Kerr black hole, highlighting the deviations between the two models. 

\begin{figure*}[htbp]
  \centering
    \includegraphics[scale = 0.48]{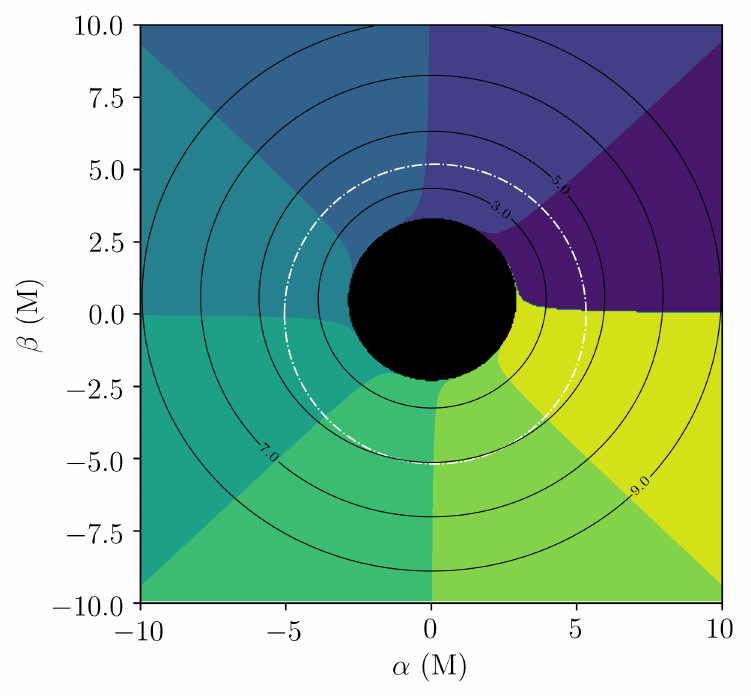}
    \includegraphics[scale = 0.48]{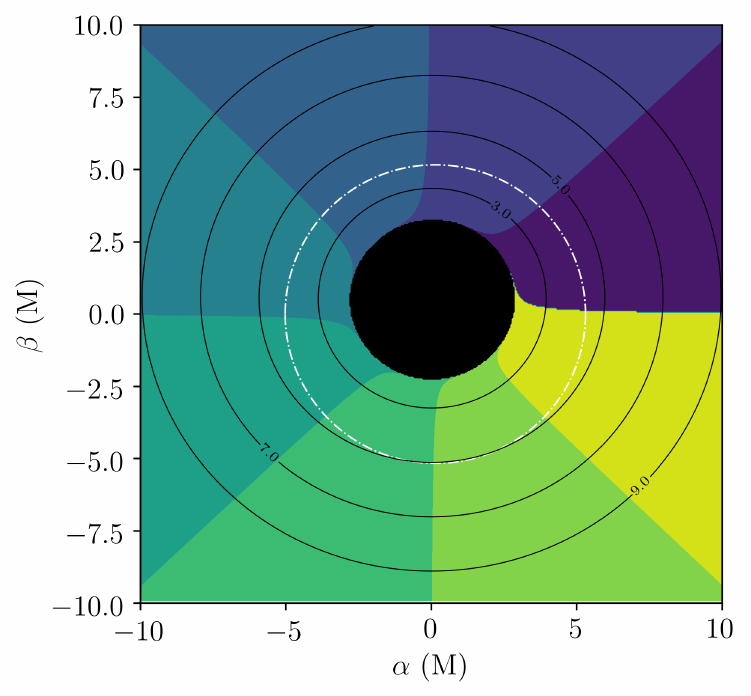}
    \includegraphics[scale = 0.48]{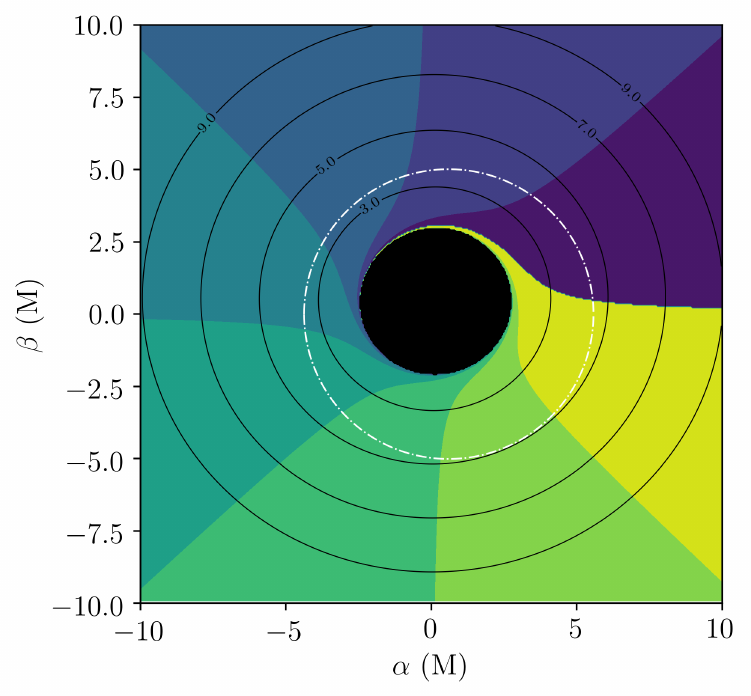}
    \includegraphics[scale = 0.48]{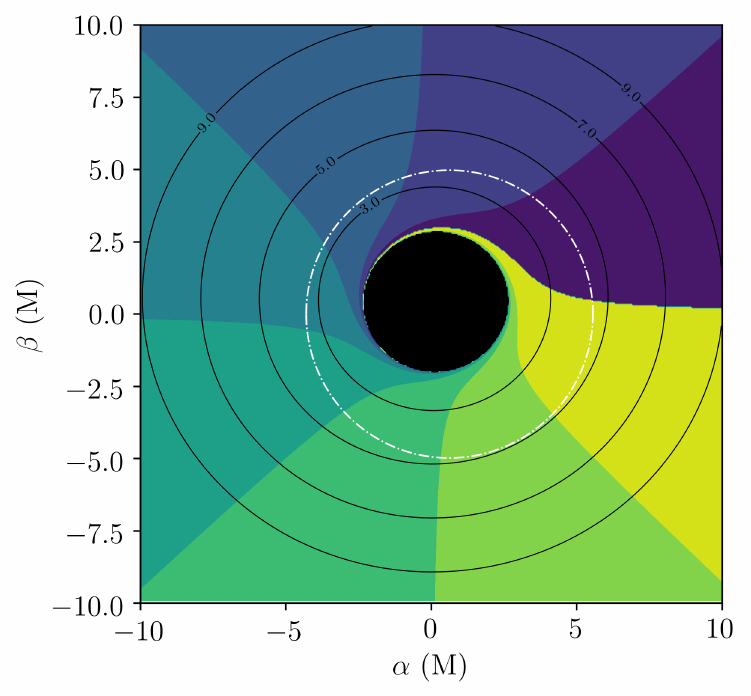}
    \includegraphics[scale = 0.48]{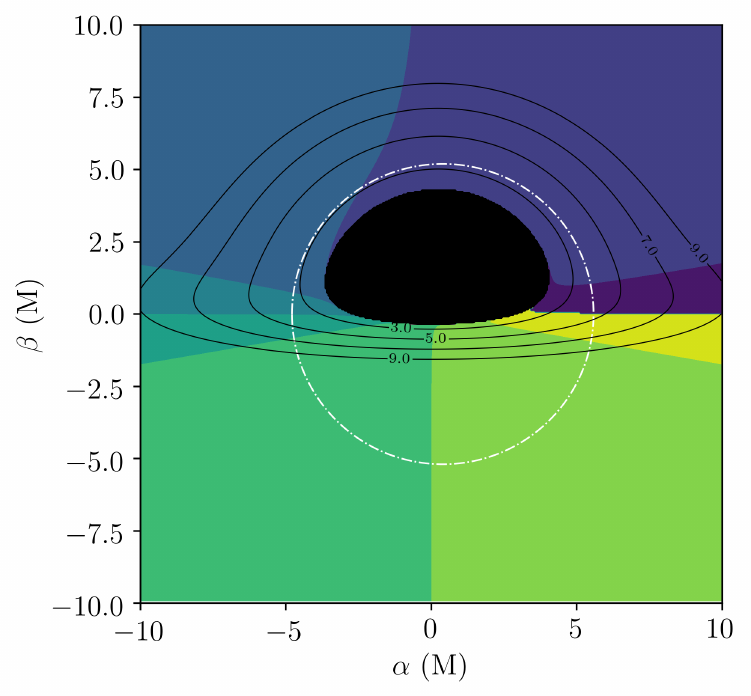}
    \includegraphics[scale = 0.48]{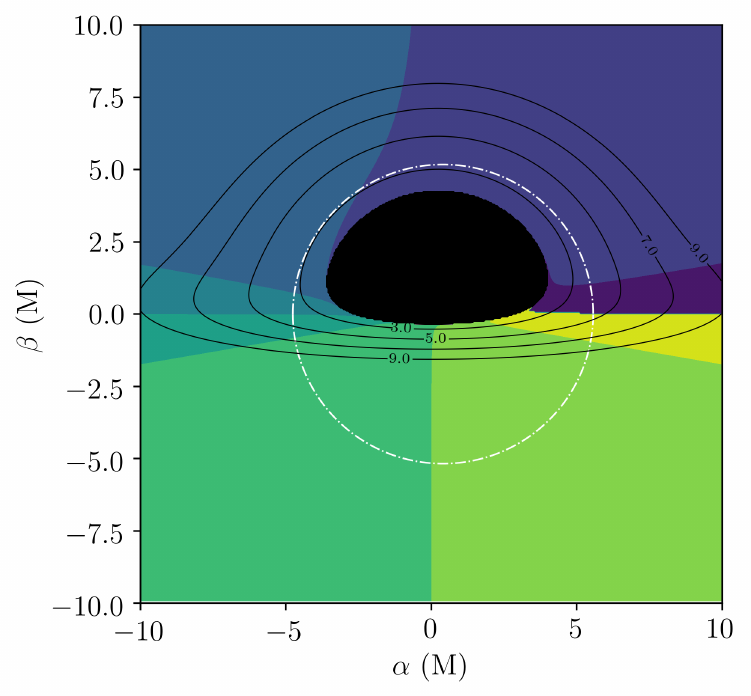}
    \includegraphics[scale = 0.48]{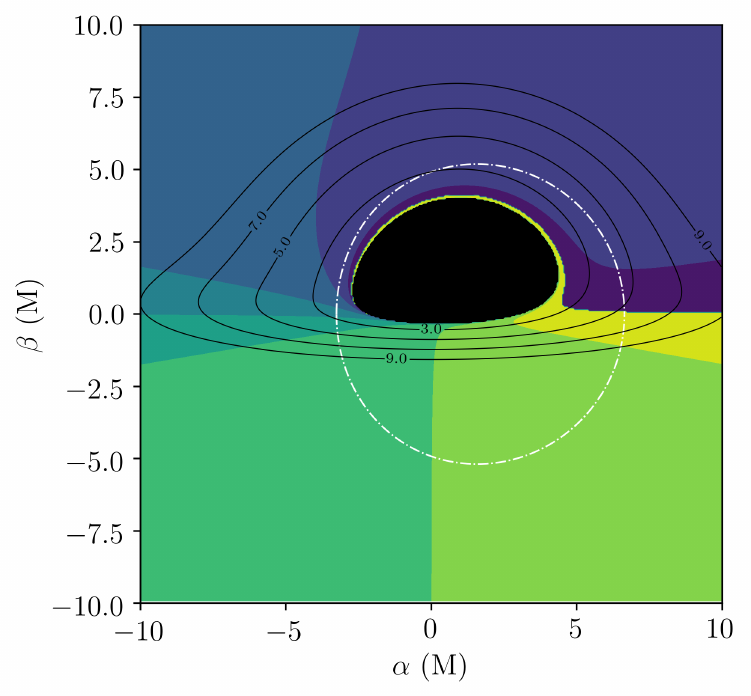}
    \includegraphics[scale = 0.48]{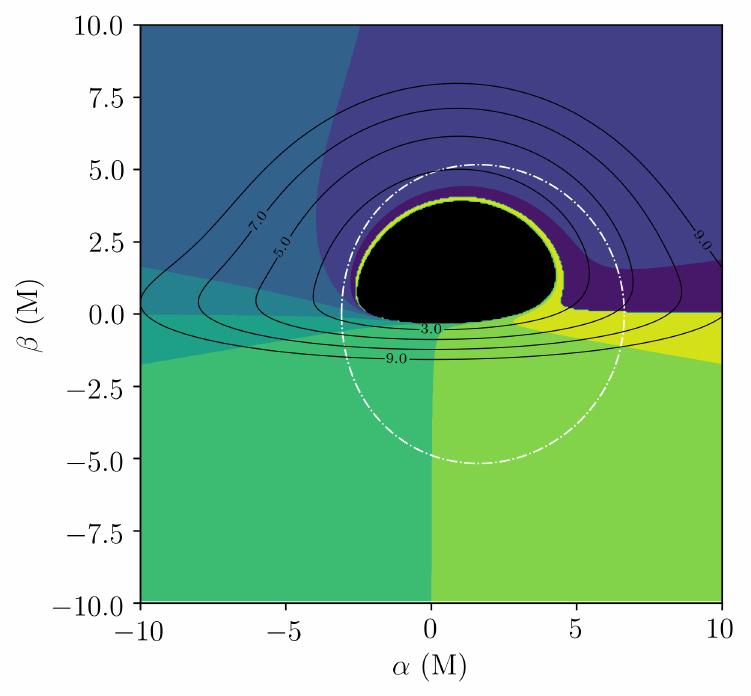}
\caption{Comparison of direct image (or $n=0$ lensing image) for both Kerr black holes (left column) and rotating hairy black holes with hairy parameters $\delta = 1, h_0=1$ (right column), including the apparent horizon (dark region), apparent source rings $r_s =3,5,7,9 M$ (black lines), as well as the critical curve (white lines). The observer is located at $r_0 = 10^4 M$. The first to fourth rows correspond to black hole spins and inclination angels with $(a=0.2, i=20^\circ), (a=0.8, i=20^\circ), (a=0.2, i=80^\circ), \text{and} (a=0.8, i=80^\circ)$, respectively. The gradient color bands represent the azimuthal angles $\phi_s$ of the thin accretion disk, which are divided equally into eight patches.}
\label{dr}
\end{figure*}

In Fig.~\ref{dr}, we present the direct image (or $n=0$ lensing image) for both Kerr black holes and rotating hairy black holes with hairy parameters $(\delta = 1, h_0=1)$. The comparison is performed for different black hole spins ($a=0.2,\,0.8$) and inclination angles ($i=20^\circ,\,80^\circ$), with the observer located at $r_0 = 10^4 M$. We also show the apparent horizon, apparent source rings ($r_s = 3, 5, 7, 9 M$), and the critical curve. The gradient color bands represent the azimuthal angles $\phi_s$ of the thin accretion disk, which are divided equally into eight patches. The left and right column shows the results of Kerr black hole and rotating hairy black hole with $(\delta = 1, h_0=1)$, respectively. 

From Fig.~\ref{dr}, we observe several common features in the direct images, which vary significantly with both the black hole spin and the inclination angles. A higher black hole spin results in a smaller apparent horizon and a more eccentric critical curve. As the spin increases, the azimuthal dragging becomes more pronounced, leading to more noticeable distortions in the image. This is due to the frame-dragging effect of rotating black holes. Additionally, larger inclination angles cause asymmetries in the apparent horizon, the source rings, and the azimuthal angles, making the overall structure more complex. When comparing the direct images of two different black holes, we observe that at lower spin values, the images are almost identical. However, as the spin increases, the differences between the two black holes become more apparent. For a given spin and inclination angle, the rotating hairy black holes exhibit a smaller apparent horizon and stronger azimuthal dragging compared to the Kerr black hole.

\begin{figure*}[htbp]
  \centering
    \includegraphics[scale = 0.55]{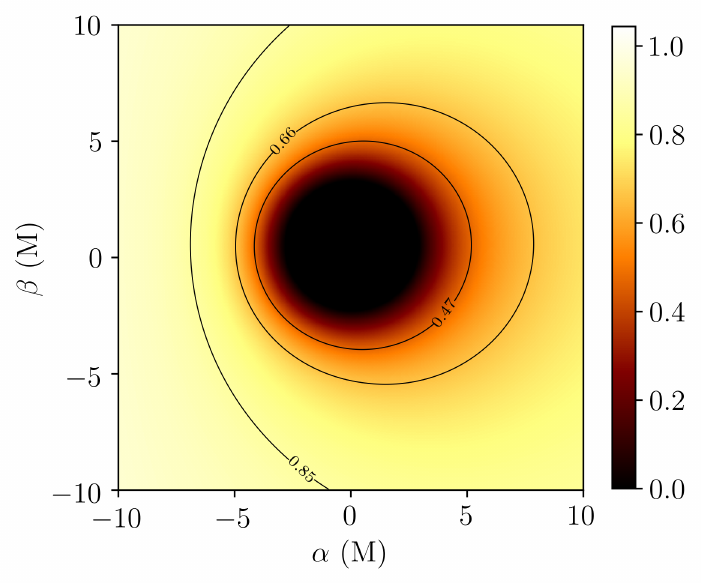}
    \includegraphics[scale = 0.55]{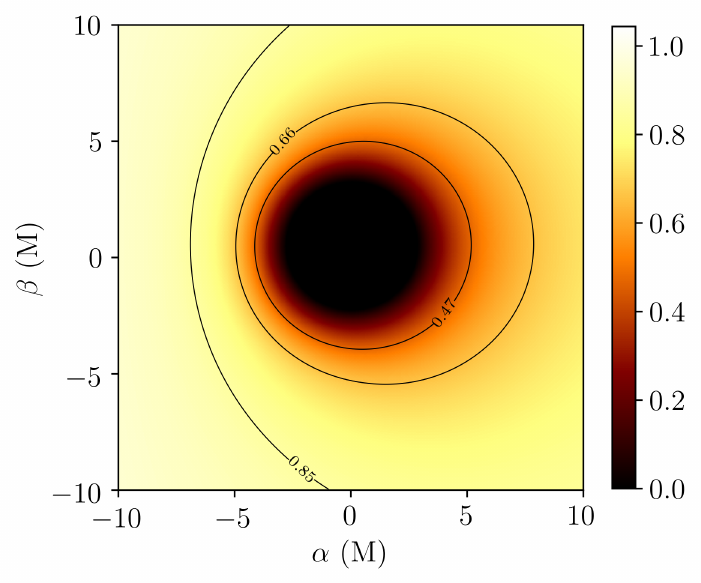}
    \includegraphics[scale = 0.55]{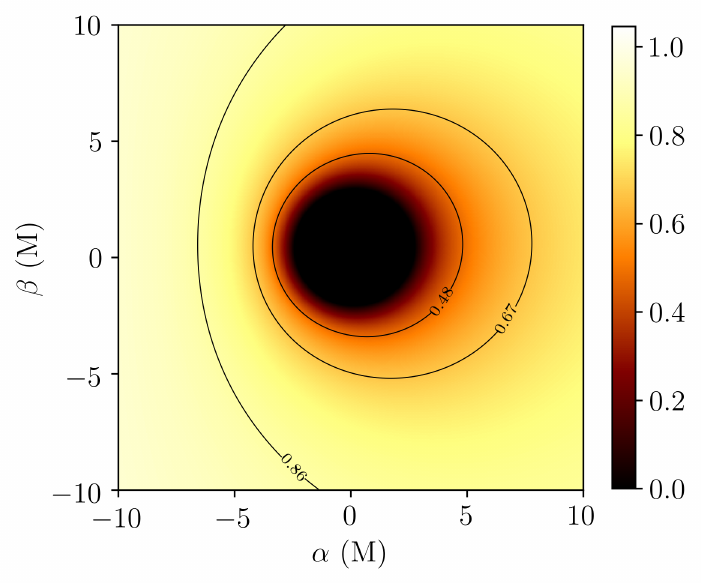}
    \includegraphics[scale = 0.55]{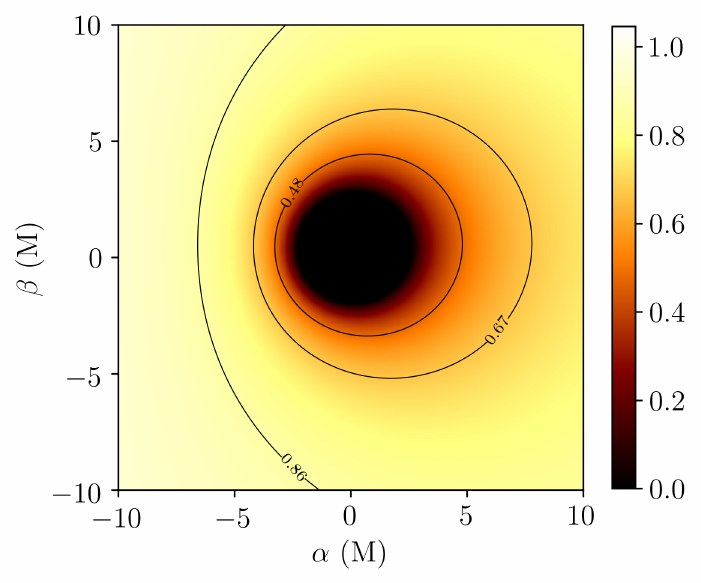}
    \includegraphics[scale = 0.55]{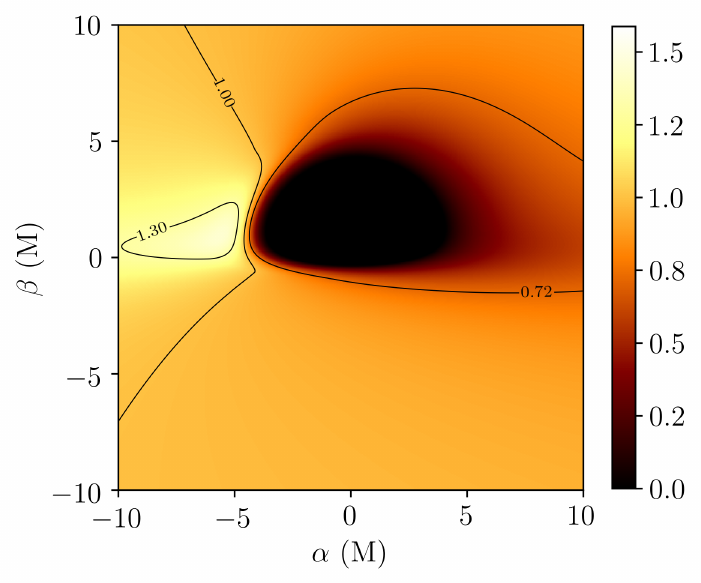}
    \includegraphics[scale = 0.55]{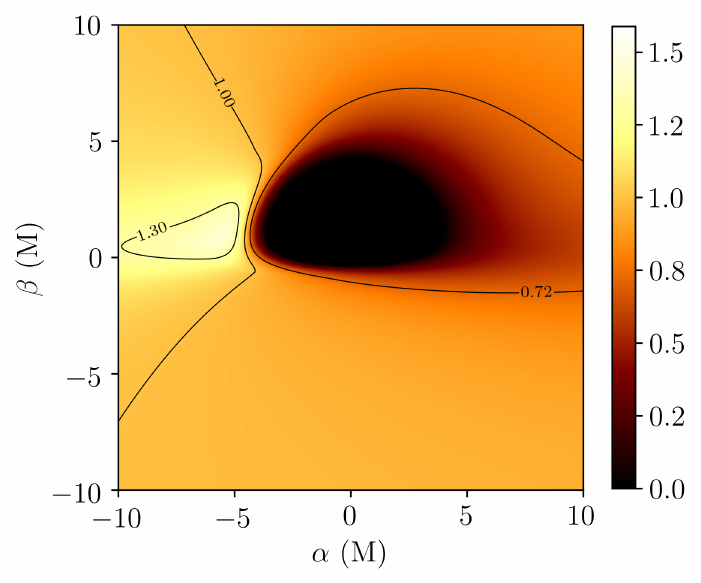}
    \includegraphics[scale = 0.55]{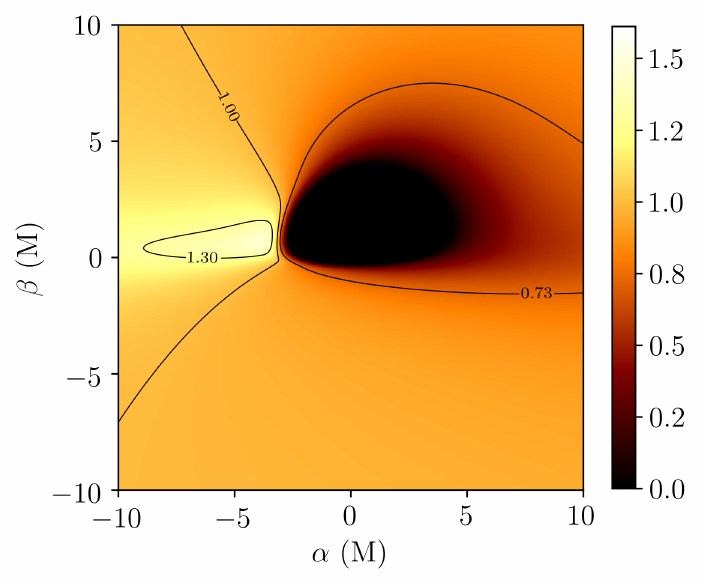}
    \includegraphics[scale = 0.55]{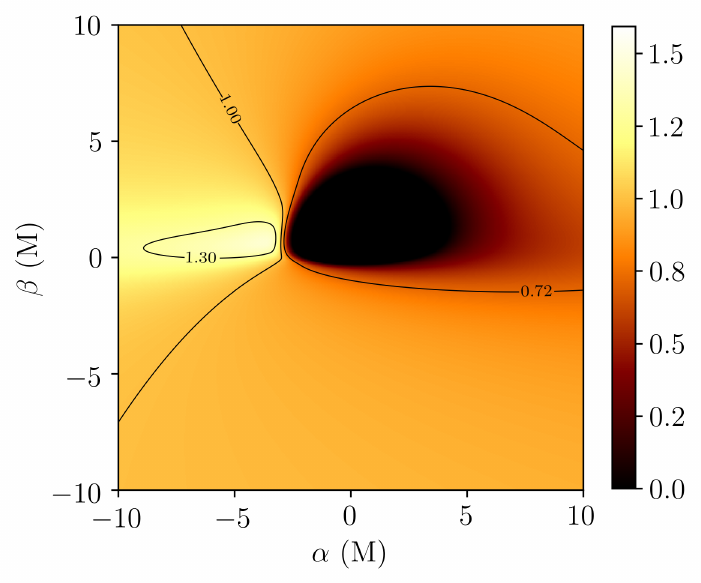}
\caption{Comparison of redshift factor on the direct image for Kerr black holes (left column) and hairy black holes with hairy parameters $\delta = 1, h_0=1$ (right column). The observer is located at $r_0 = 10^4 M$. The first to fourth rows correspond to black hole spins and inclination angels with $(a=0.2, i=20^\circ), (a=0.8, i=20^\circ), (a=0.2, i=80^\circ), \text{and}\, (a=0.8, i=80^\circ)$, respectively. The contours show the same values of redshift factor crossing the whole image plane, and the color bars in each subplot represent the redshift factor values.}
\label{gfimage}
\end{figure*}

Next, we investigate the redshift distribution on the direct image. In Fig.~\ref{gfimage}, we show the redshift factor $\chi$ (defined in Eq.~\ref{red}) with different black hole spins ($a=0.2,\,0.8$) and inclination angles ($i=20^\circ,\,80^\circ$). The observer is located at $r_0 = 10^4 M$. Since $\chi = 1/(1+z)$, we can interpret the value of $\chi$ as follows: For $\chi = 1$, there is no redshift, which corresponds $z = 0$. When $\chi < 1$, the photons are redshifted, indicating that the wavelength of the photons has increased. On the other hand, when $\chi > 1$, the photons are blue-shifted.

From Fig.~\ref{gfimage}, we can observe that the black hole spin only has a minor effect on the redshift distribution, while the inclination angle significantly influences the observed redshift. In particular, a photon can exhibit a noticeable blueshift for certain region when observed at large inclination angles. As the photon approaches the black hole horizon, the redshift increases dramatically. Inside the apparent horizon, the redshift becomes infinite. Furthermore, the differences between the Kerr black hole and the rotating hairy black hole become more pronounced at higher spins, whereas they appear nearly identical at lower spin values. This again indicates that the spin plays a crucial role in distinguishing the observational characteristics of these two types of black holes.

\begin{figure*}[htbp]
  \centering
    \includegraphics[scale = 0.55]{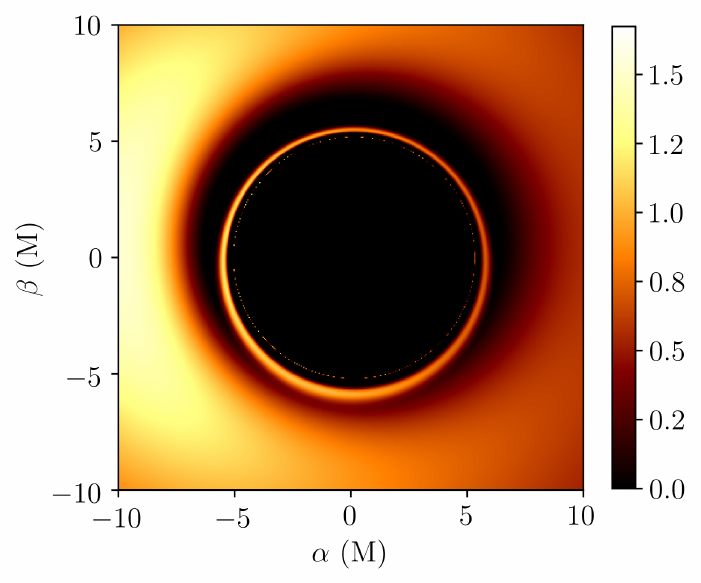}
    \includegraphics[scale = 0.55]{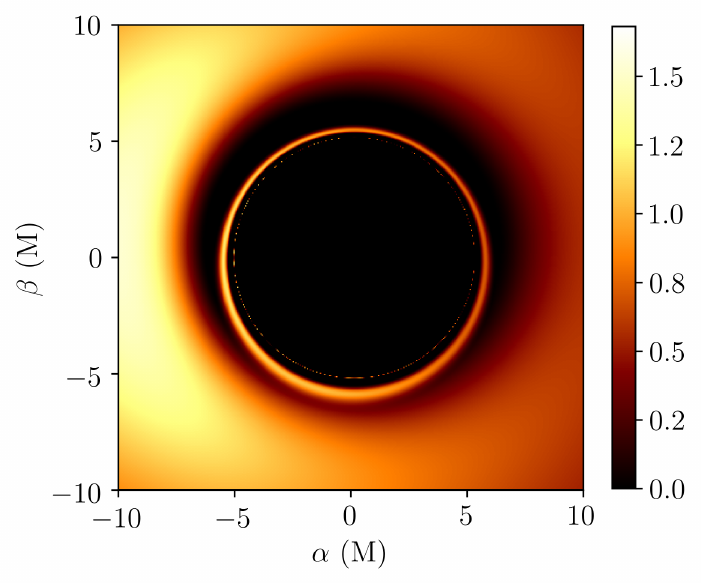}
    \includegraphics[scale = 0.55]{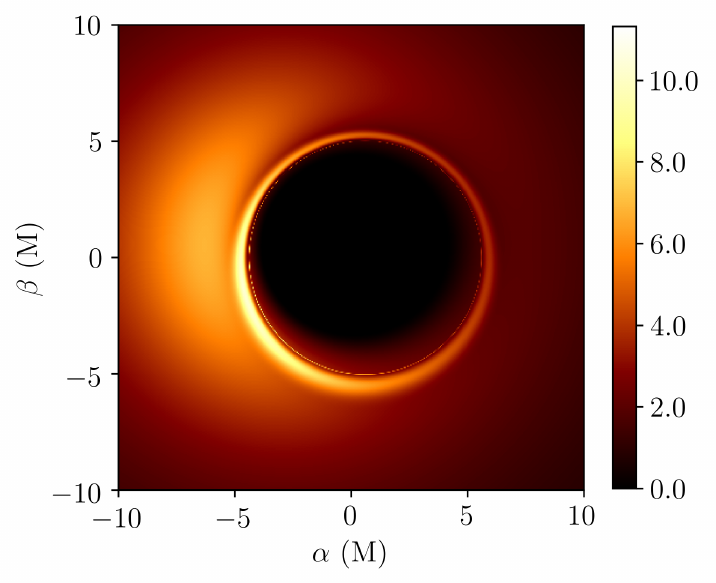}
    \includegraphics[scale = 0.55]{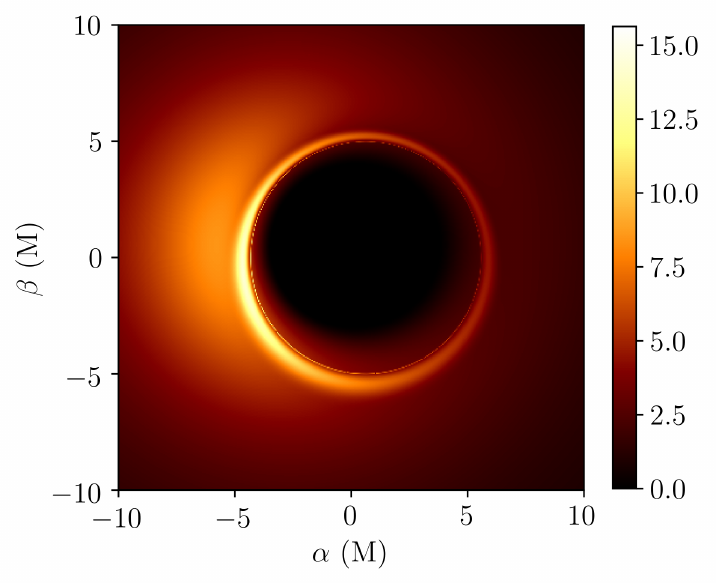}
    \includegraphics[scale = 0.55]{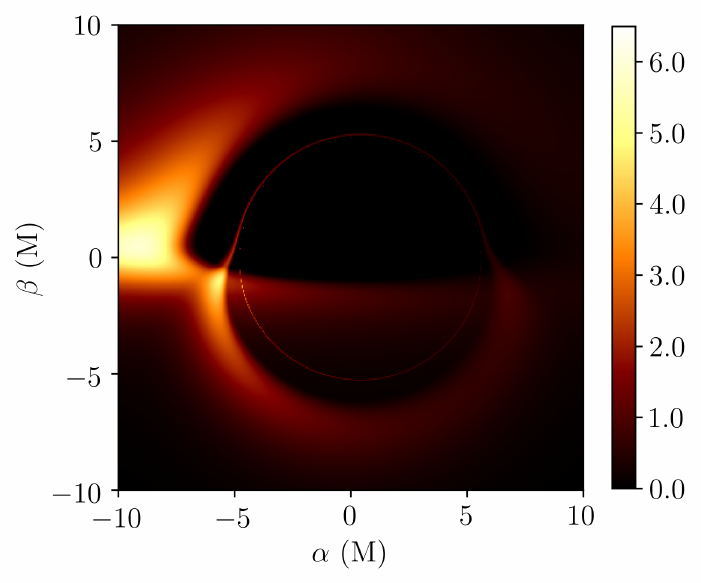}
    \includegraphics[scale = 0.55]{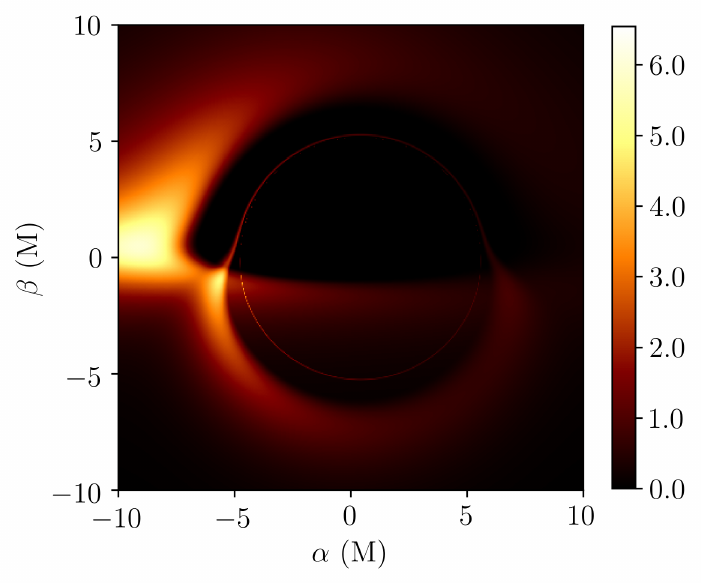}
    \includegraphics[scale = 0.55]{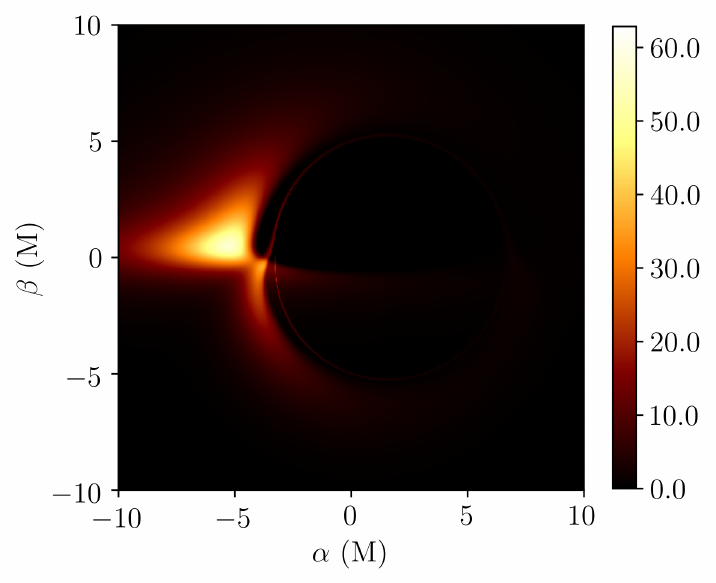}
    \includegraphics[scale = 0.55]{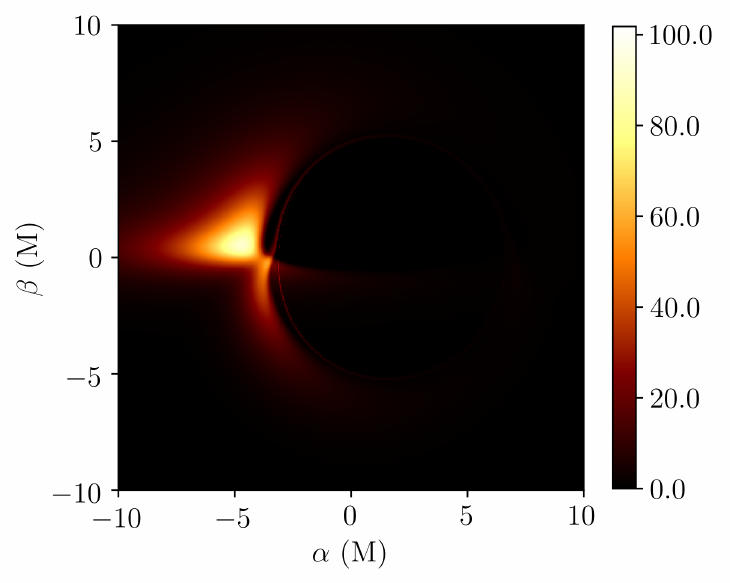}
\caption{Comparison of black hole images for Kerr black holes (left column) and rotating hairy black holes with $(\delta = 1, h_0=1)$ (right column). The observer is located at $r_0 = 10^4 M$. The first to fourth rows correspond to black hole spins and inclination angels with $(a=0.2, i=20^\circ), (a=0.8, i=20^\circ), (a=0.2, i=80^\circ), \text{and}\, (a=0.8, i=80^\circ)$, respectively. The color bars in each subplot represent the brightness of the black hole image, with brighter regions corresponding to higher bolometric values. The source flux profile $\mathcal{F}_s(r)$ for generating the observed images are taken from Fig.~\ref{f} by making $\dot m =1$ and multiplying the values with a factor of $10^5$.}
\label{bhimage}
\end{figure*}

The two figures presented above,  Fig.~\ref{dr} and Fig.~\ref{gfimage}, serve as a foundation for analyzing the bolometric image of thin accretion disk around rotating hairy black holes. By utilizing these results, we are able to explore its optical appearances and compare it with the Kerr black hole. Fig.~\ref{bhimage} presents a comparison of the bolometric images of the Kerr black hole and the rotating hairy black hole with $(\delta = 1, h_0 = 1)$. The images are shown for different black hole spins ($a=0.2,\,0.8$) and inclination angles ($i=20^\circ,\,80^\circ$). The observer is at $r_0 = 10^4 M$. The radiative flux profile $\mathcal{F}_s(r)$, used to generate the observed bolometric images, is taken from Figure~\ref{f}, with $\dot m = 1$, and the flux values are scaled by a factor of $10^5$.

From Fig.~\ref{bhimage}, it is clear that the spin of the black hole plays a crucial role in distinguishing between the Kerr black hole and the rotating hairy black hole. For lower values of spin, the images of both black holes appear nearly indistinguishable. However, as the spin increases, the rotating hairy black hole exhibits a significantly higher brightness in the bolometric images compared to the Kerr black hole. This difference in brightness can be attributed to the more pronounced emission from the rotating hairy black hole, as shown in Fig.~\ref{f}. Furthermore, for a given spin and inclination angle, the rotating hairy black hole displays a more compact structure in its image. This suggests that the rotating hairy black hole, due to its modifications to the black hole spacetime, produces distinctive observational signatures that can be different from those of Kerr black holes.

\section{Conclusion}\label{sec5}

Rotating hairy black holes, derived using the gravitational decoupling method, represent a well-motivated extension of general relativity by incorporating the effects of unknown fields or matter near black holes. In this study, we analyze the radiative properties and optical appearance of thin accretion disks surrounding these black holes. Specifically, we examine the radiative flux, temperature distribution, differential luminosity, and bolometric images of the disks.

Our numerical results demonstrate significant deviations from the Kerr black hole predictions, particularly for rapidly rotating black holes or in the inner regions of the accretion disk. These deviations suggest that the rotating hairy black hole introduces distinct physical effects that enhance the radiative efficiency of the accretion disk and increase the brightness of black hole images. Such enhancements are particularly notable in the energy output and observable luminosity profiles, providing key insights into the potential detectability of these modified black holes.

Enhanced radiative signatures and distinct image characteristics may offer new opportunities to probe the physics of black holes beyond the GR paradigm. The findings presented here are expected to serve as a theoretical basis for interpreting upcoming high-precision observational data, particularly from next-generation telescopes and interferometers. This research contributes to a deeper understanding of how modifications to general relativity manifest in observable phenomena around compact astrophysical objects.

During the preparation of this manuscript, we became aware of a related study conducted by Xiao-Mei Kuang and her collaborators (coming soon), which also investigates the images of the rotating hairy black hole discussed in this paper. While our work aim to study the optical appearance of thin accretion disk, they focus on the strong gravitational lensing effects of rotating hairy black hole. Both works contribute valuable insights to the observational features of rotating hairy black hole, highlighting the importance of continued exploration in this field.

\section*{Acknowledgments}
Z. Li gratefully acknowledges the financial support from the Start-up Funds for Doctoral Talents from Jiangsu University of Science and Technology. X.-K. Guo is supported by Yancheng Institute of Technology (xjr2024030).
\\
\\

\end{document}